# Healthcare

Shoumen Datta

Julian Goldman

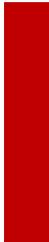

*This is the true joy in life, the being used for a purpose you consider a mighty one, the being a force of nature rather than a feverish, selfish clod of ailments and grievances complaining that the world will not devote itself to making you happy.*

George Bernard Shaw

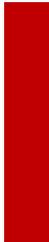

# Digital Transformation of the Healthcare Value Chain: Emergence of Medical Internet of Things (MIoT) may need an Integrated Clinical Environment, ICE Platform

Table of Contents

# Digital Transformation of the Healthcare Value Chain: Emergence of Medical Internet of Things (MIoT) may need an Integrated Clinical Environment, ICE Platform

Shoumen Palit Austin Datta, PhD
Senior Member, MIT Auto-ID Labs and Research Affiliate, Dept of Mechanical Engineering
Massachusetts Institute of Technology
Senior Scientist, Medical Interoperability Program, MDPnP Labs
Massachusetts General Hospital, Harvard Medical School
shoumen@mit.edu or sdatta8@mgh.harvard.edu (additional info http://bit.ly/IOT-MIT)

Julian M Goldman, MD
Medical Director, Biomedical Engineering, Department of Anesthesiology
Director, Medical Interoperability Program, MDPnP Labs
Partners HealthCare System, Massachusetts General Hospital, Harvard Medical School
jmgoldman@mgh.harvard.edu (additional info http://www.juliangoldman.info/)

## ABSTRACT

The complexity of the healthcare ecosystem and the trans-disciplinary convergence which is essential for its function, makes it difficult to address healthcare as one domain. Divided in various stages for ease of management, healthcare is quintessentially a continuum which commences at birth and may cease at death. In this paper, we have attempted to take a broad view but it still isn't broad enough. We have focused on issues, such as, data and interoperability, yet it scratches only the surface of these sub-domains. For the specific problem at hand, we propose an "http" approach to adopt open standards to enable medical equipment and devices to share and synthesize data using ICE (integrated clinical environment), a medical device interoperability platform. ICE may help reduce deaths due to preventable medical errors. Thinking about the future of data in personal healthcare, we propose creating an ecosystem which contributes data from multiple health related domains. Data curation and analysis of the information may boost our health related knowledge. Increasing connectivity and improving infrastructure may help, among other things, to uncover facts and observations which may influence the future of global health.

**BACKGROUND** (US-CENTRIC)

The legendary editor of the august New England Journal of Medicine, Arnold S Relman (1923-2014), offered an incisive perspective when he delivered the 171st Annual Discourse at the Massachusetts Medical Society on May 21, 1980 [1].

"According to an article in the *Wall Street Journal* of December 27, 1979, the net earnings of health-care corporations with public stock shares rose by 30 to 35 per cent in 1979 and are expected to increase another 20 to 25 per cent in 1980. A vice-president of Merrill Lynch appeared a few months ago (prior to this article) on "Wall Street Week," the public television program, to describe the attractions of health-care stocks. According to this authority, health care is now the basis of a huge private industry, which is growing rapidly, has a bright future, and is relatively invulnerable to recession. He predicted that the health business would soon capture a large share of the health-care market and said that the only major risk to investors was the threat of greater government control through the enactment of comprehensive national health insurance or through other forms of federal regulation" [1].

The sluggish progress of the US national health insurance initiatives [2] and the pugilistic powers in action [3] are indicative of the pugnacious influence of private health-care industries on national health policy. A broad national health-insurance program, with the inevitable federal regulation of costs, is an anathema to the medical-industrial complex, just as a national or global disarmament policy may be to the military-industrial complex. President Eisenhower was concerned about the unhealthy aspirations of the military-industrial complex and emphasized "*we must guard against the acquisition of unwarranted influence*" in his farewell address [4]. A similar admonition is applicable to the healthcare industry. Today's medical-industrial complex is replete with aggressive enterprises vying for global dominance of epic proportions and maximizing profit taking at every step.

The transmutation of healthcare, once guided by the ethos of the Hippocratic Oath [5], to a commodity guided by the free market economy to improve efficiency and quality, is deeply flawed. Application of business values [6] and operational optimizations are uninformed efforts by well-intentioned bean counters. Patients are not "consumers" or "clients" and doctors are not "partners" or "service providers" in the classical context of Adam Smith [7]. However, business relationships and "best practices" may be applicable when a hospital or a clinic is purchasing bathroom tissue or contracting for janitorial services or outsourcing valet parking. The latter is an operational function but cannot be referred to as healthcare.

Hence, tools of operations management, for example, supply chain management, are useful only to a limited extent to things, equipment, objects or facilities, which are very likely to be less uncertain in their use or application because the goods or inventory, usually, are uninfluenced or unrelated to the practice of healthcare. These observations, however, have had little impact on the vast cost of consulting fees extracted from the healthcare industry by major firms [8] who continue to pontificate about business best practices to healthcare.

Patients who are sick, or worried that they may be sick, generally, are neither capable of understanding their physiological status nor inclined to shop around for bargains. The value of life often far outweighs the consideration of cost. Patients and their families seek and demand the best care they can get irrespective of the price. Hence, the classic laws of supply and demand are ill suited because healthcare "consumers" may not subscribe to the usual incentives to be prudent, discriminating and frugal in their decision. This "decision" is not about profit and loss, even if glib management consultants may view it as a *purchase*.

Health care is neither synonymous nor interchangeable with "medical marketplace" because the canonical rules of competitive economic equilibrium [9] are unlikely to be applicable. The tireless pursuit of the for-profit business process consultants to inflict the so-called supply chain principles from the "marketplace" on to healthcare is fraught with problems. It exhibits a flagrant disregard for the foundational distinction that must choose value over cost in the practice of medicine or the effort necessary to save even a single life. Economies of scale or risk pooling are not always applicable or even desired in health care.

The root of the disequilibrium in healthcare is the heavy, often total, dependence of the patient (irrationally referred to as the consumer) on the medical practitioner (nurse or physician, viewed in the business framework as a service provider).

The business consultants, administrators and software packages peddling the operational principles extracted from retail stores or grocery chains or manufacturing plants may not be fully cognizant about the seminal work by Kenneth Arrow referred to as information inequality [10]. The latter catalyzed more in depth analysis of markets with asymmetric information pioneered by George A. Akerlof, A. Michael Spence and Joseph E. Stiglitz [11].

It is interesting to observe that few business students, middle managers and corporate executives are able to connect the fact that the Forrester Effect [12] which later morphed as the Bullwhip Effect [13] is partly due to information asymmetry [14]. The resurgence of RFID [15] to ignite the digital supply chain [16] has had limited impact on the reduction of information asymmetry. The reasons may include dead weight of old technology, lack of an engineering systems approach [17], inability to evolve out of the organizational "silo" frameworks and punctuated connectivity. Taken together, productivity gains, if used as a (key) performance indicator (KPI), remains a paradox [18] while limits on our data driven prediction abilities [19] makes predictive analytics more of an art and less of a paradigm.

The incisive foresight available from the application of the principles of information asymmetry to healthcare may explain why the usual assumptions about competitive free markets do not apply. There aren't any known mechanism to the bridge the chasm of medical knowledge between the patient and physician in order to generate "equilibrium" through the establishment of "symmetric" information. The patient does not choose the plethora of medical tests or the regimen of procedures or plan for medication. Physicians decide the course of action. Hence, it is the physician who will influence 70% or 80% of all the expenditures associated with healthcare.

The potential for financial abuse, therefore, is obvious, when private for-profit companies enter the market. Private healthcare companies can conspire to influence the decisions of the physicians to maximize profit [20]. The physician is a profit center – actively engaged in the business of the *medical marketplace* which is a service industry. This scenario may fit the corner service station at the intersection of Happy and Healthy, serving gas on demand.

In Goldfarb vs Virginia State Bar, the US Supreme Court [21] handed down a landmark decision that found that the business activities of professionals were properly subject to antitrust law. Today, we are dealing with increasingly caustic consequences of that decision, as astutely pointed out by Late Arnold S Relman [22] in his 101st Shattuck Lecture at the Annual Meeting of the Massachusetts Medical Society in Boston on May 18, 1991 [23].

It would be a heresy to temporarily conclude this sketchy US-centric background without mentioning the influence of physicians and the organizations they champion (for example, the almighty AMA or America Medical Association). The forbidding political landscape in US healthcare reform was shaped, in part, by the physicians and their refusal to include government health insurance in the 1935 Social Security Act. In 1945, the AMA lobbied against President Harry Truman's proposed universal health insurance program and delivered the fatal blow. In 1965, much to the chagrin of the AMA, it met with partial defeat when President Lyndon Baines Johnson created Medicare, a federal health insurance program for the elderly, and Medicaid, a combined federal-state program for the poor. President Bill Clinton's failure was, in part, due to the opposition party who may have bullied the US Chamber of Commerce into withdrawing its support, as published by a senior member of the healthcare reform team, Paul Starr of Princeton University [24].

The timbre of compromises, tempered success and the catastrophic victory of President Obama's Affordable Care Act on 23 March 2010 [25] capped a struggle for US healthcare reform that commenced even before 1935. The legal name of "Obamacare" is The Patient Protection and Affordable Care Act. One cornerstone of PPACA is patient protection. This paper and our mission is focused on patient safety. Without patient safety, it may be well-nigh impossible to adequately address what is necessary to ensure patient protection.

Quote from: *The Health Care Industry: Where Is It Taking Us?* by A. S. Relman [23]

Physicians can no longer act collectively on matters affecting the economics of practice, whether their intent is to protect the public or simply to defend the interests of the profession. Advertising and marketing by individual physicians, groups of physicians, or medical facilities, which used to be regarded as unethical and were proscribed by organized medicine, are now protected - indeed, encouraged - by the Federal Trade Commission.

Advertisements now commonly extol the services of individual physicians or of hospital and ambulatory facilities staffed by physicians. Most of them go far beyond simply informing the public about the availability of medical services. Using the slick marketing techniques more appropriate for consumer goods, they lure, coax, and sometimes even frighten the public into using the services advertised.

I recently saw a particularly egregious example of this kind of advertising in the *Los Angeles Times.* A freestanding imaging center in southern California was urging the public to come for magnetic resonance imaging (MRI) studies in its new "open air" imager, without even suggesting the need for previous examination or referral by a physician. The advertisement listed a wide variety of common ailments about which the MRI scan might provide useful information — a stratagem calculated to attract large numbers of worried patients whose insurance coverage would pay the substantial fee for a test that was probably not indicated.

Many respectable institutions and reputable practitioners advertise in order to bring their services to the public's attention. But in medical advertising there is a fine line between informing and promoting; as competition grows, this line blurs. Increasingly, physicians and hospitals are using marketing and public relations techniques that can only be described as crassly commercial in appearance and intent.

Before it was placed under the protection of antitrust law, such advertising would have been discouraged by the American Medical Association (AMA) and viewed with disfavor by the vast majority of physicians. Now it is ubiquitous, on television and radio, on billboards, and in the popular print media. Of course, not all medical advertising is as sleazy.

## IN PRAISE OF OTHER IMPERFECTIONS

Four common [26] approaches are (a) private-sector (USA, Switzerland), (b) national health service (UK), (c) provincial government health insurance (Canada) and (d) social insurance (France, Netherlands, Germany). Founded in 1883, by Otto von Bismarck, Germany's Statutory Health Insurance (SHI) program insures about 99.9% of the country's population and is administered by private not-for-profit organizations, authorized by law to wield power on behalf of payers and providers. The federal government intervenes in the interest of public goods if the broader interests of society are neglected [27]. The SHI Health Care Structures Reform Law, passed in 2012, created a third care sector called integrated ambulatory specialist care [28]. The latter (IASC) and other efforts (in 2004) are an attempt to mitigate the negative impact on integration due to previous segregation of ambulatory and hospital care delivery. The segregation dates back to a 1931 decree by Chancellor Heinrich Brüning, which granted physicians in private practice a monopoly on ambulatory care and essentially prohibited hospitals from providing outpatient care (the emergency decree was a direct result of a physician strike).

Economist Robert J Evans posits that the ethos of equality in the provincial healthcare system of Canada reflects a principle similar to equality before law [29]. The Canadian system is an imperfect hybrid of public funding (government taxes), private providers and universal coverage which is comprehensive in scope, affordable, single-payer, provincially administered yet works as a "national" healthcare system. It reflects political compromises, in face of powerful opposition [30].

The lofty idealism in the Canadian system is but a poisoned chalice to the US medical-industrial complex. Big pharma, manufacturers and insurance companies poured more than half a billion US dollars to lobby US politicians [31] and even more was invested to commission studies with "alternative facts" to advocate that a US national health insurance scheme modeled on the Canadian system would not work. An excess of a quarter billion US dollars went into the coffers of political parties [32] to convince Americans that they will get inferior health care coverage and fewer choices with a Canadian system [33].

The US political momentum toward the lowest possible common denominator is a tool to preserve corporate interests in healthcare, for example, the lucrative dialysis business [34]. Measures and medication to reduce the need, cost and deaths due to dialysis [35] are often subjected to the vagaries of obfuscation or appears to be mired in chronic controversies [36] orchestrated with Machiavellian shrewdness. Public illiteracy of science and medicine goads forward profit optimization routines. Those who know and can shed light on these unscrupulous practices are failing to sustain their opprobrium and silence fuels repetition.

## INTRODUCTION

The ills of the US medical-industrial complex are not unique to the US. Many of the medical equipment manufacturers [37] are global. The product-driven manufacturing business model prevalent in the equipment and device sector is nuclear. The product is supposed to spell profit for companies and any attempt to share parts, feature or information about the equipment or device is viewed as detrimental for profitability. This is a general observation and not a specific behavior exclusive to the US medical-industrial complex.

This is also where medical manufacturing must distinguish itself from other industry verticals. The equipment and devices used in medical and healthcare system must be optimized for performance which, generally, may focus on one physiological function and can be used on one individual or one patient, at a time. A heart rate monitor cannot multi-task and monitor Jane and Joe, simultaneously. Hence, the principles of business (best practices, risk pooling, supply chain, economies of scale and similar operational lessons) from most other domains are relatively sterile, since the end point of the activity for the medical manufacturer is the well-being of an animal or one human being, at a time.

The human being is not the patellar disc, myocardial infarction or glomerular nephritis. A human being may have one or more of those symptoms, syndromes of afflictions but the entire person or the "whole" patient must be the focus of health care (albeit, there may exist different stages of acuity, which may or may not be applicable in this scenario).

Safety is of paramount importance. Safety is about the status of the patient, as a whole.

Surgical wizardry may save the life of a person involved in an automobile collision. Yet the post-operative patient may succumb to patient-controlled over-dose of analgesia in an ICU.

These deaths due to errors are often due to lack of integration of patient safety processes at the level of the medical devices, several of which may be attached to monitor each patient.

Each device or equipment is usually dedicated to monitor one function and can be used for one patient at a time. The physiological functions, essential for monitoring, make up a long list and each function may require a specific device manufactured by a specific corporation. Several different corporations are supplying these devices and equipment. The functional integration of data from devices which can reflect the physiological "status" of the "whole" patient at the point of care (nurse, physician) is not a competency within the domain of device manufacturers. OEM's are not experts in physiological status integration. Hence, perhaps, it is unfair to expect device and equipment manufacturers to bear the moral burden of integration, necessary for physiological status.

The physiological "sum" of data, from different devices, taken together, can make a significant difference between saving a life or death.

Errors leading to death of patients include errors arising out of lack of patient safety processes. A key element of patient safety is related to unintegrated device-level data. Devices are designed by manufacturers to present a pre-set, time-dependent, single function data point but usually lack integration with other single function monitors. This data segregation introduces (often, fatal) medical errors by preventing physiological status updates about the patient, as a whole, in a hospital or nursing facility or in private homes.

In the US, deaths due to medical errors (Figure 1) are the third leading cause of death [38]. That is equivalent to one or two passenger-filled jumbo jets (B747) crashing every day.

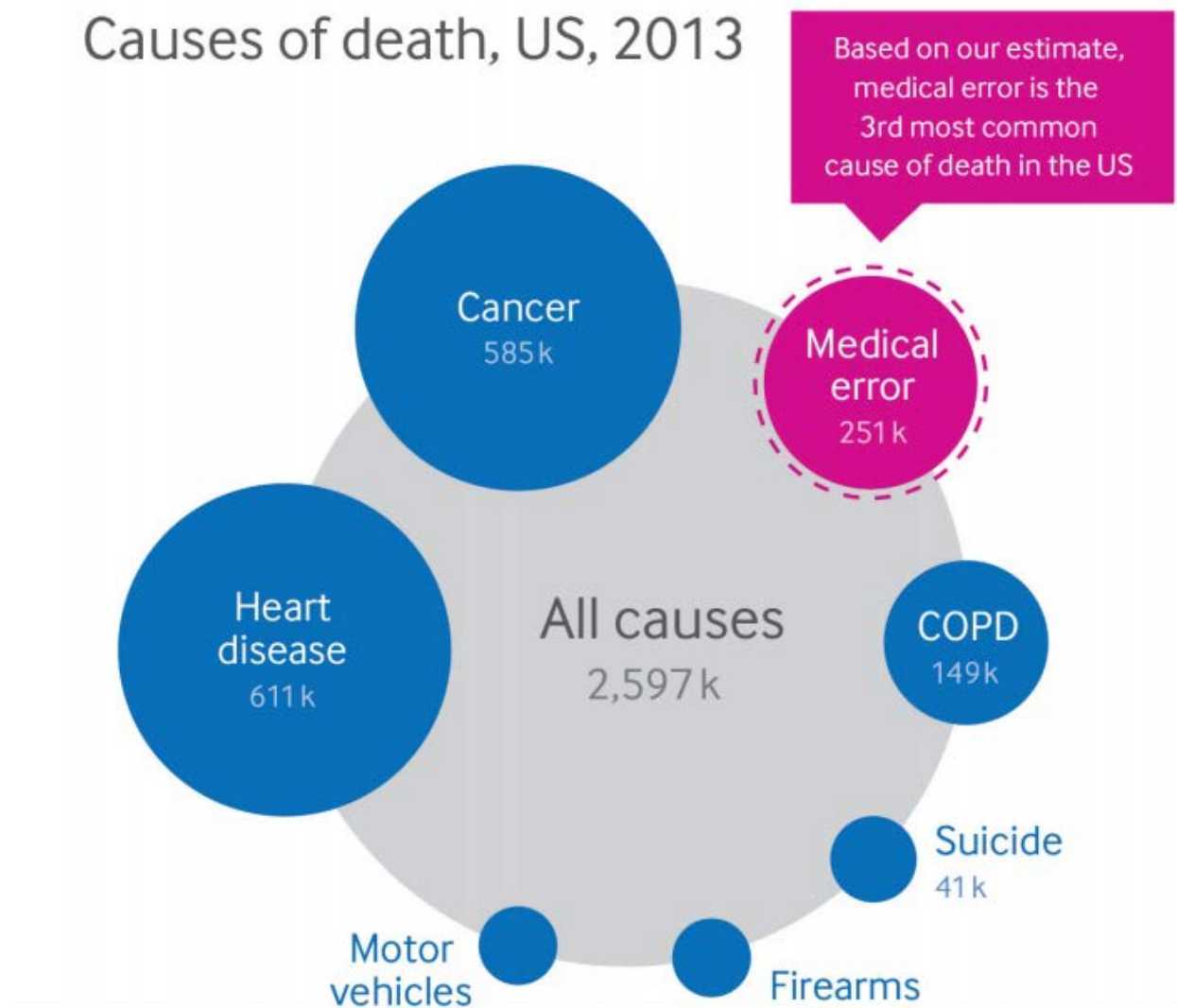

Figure 1: An estimated 220,000 – 440,000 patients die due to medical errors in the US [38]

Deaths due to medical errors appear to be less documented in OECD nations. The relative virtues of the private Swiss system may be due to its demographics (population 8 million) and economics, 4th highest GDP in the world [39]. UK National Health Service data about adverse incidents is shown in Figure 2 [40]. Grouped as accidents under the leading causes of deaths in different geographies [41], deaths due to medical errors are unclear [42].

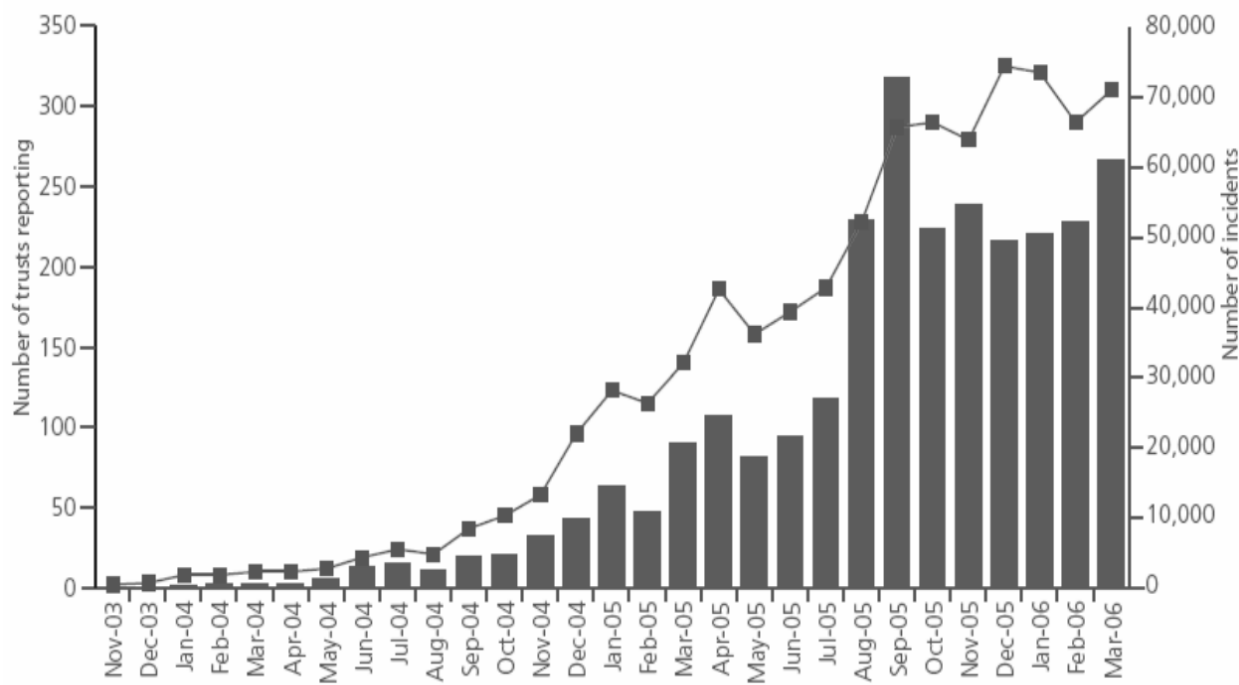

Figure 2: UK NHS Reporting Trusts and Number of Incidents (bars) in the National [40] Reporting and Learning System. NRLS is a database where adverse incidents are reported.

| Rank | Israel | EU-15 countries | USA | Canada |
|------|--------|-----------------|-----|--------|
| 1 | Malignant neoplasms | Malignant neoplasms | Heart disease | Malignant neoplasms |
| 2 | Heart disease | Heart disease | Malignant neoplasms | Heart disease |
| 3 | Cerebrovascular disease | Cerebrovascular disease | CLRD | Dementia |
| 4 | Diabetes | CLRD | Dementia | Cerebrovascular disease |
| 5 | Septicemia | Dementia | Cerebrovascular disease | CLRD |
| 6 | Kidney disease | Accidents | Accidents | Accidents |
| 7 | CLRD | Pneumonia & Influenza | Alzheimer's disease | Diabetes |
| 8 | Dementia | Diabetes | Diabetes | Alzheimer's disease |
| 9 | Pneumonia & Influenza | Alzheimer's disease | Pneumonia & Influenza | Pneumonia & Influenza |
| 10 | Accidents | Kidney disease | Kidney disease | Kidney disease |

Table 1: Leading Causes of Death as published by the Ministry of Health, Israel [41]

**PROBLEM SPACE – Preventable Fatalities due to lack of an Integrated Clinical Environment**

A patient had a laparoscopic cholecystectomy [gall bladder removal] performed under general anesthesia. At the surgeon's request, a plain film x-ray was taken during a cholangiogram [bile duct x-ray]. The anesthesiologist stopped the ventilator for the film. The x-ray technician was unable to remove the film because of its position beneath the table. The anesthesiologist attempted to help but found it difficult because the gears on the table were jammed. Finally, the x-ray was removed and the procedure recommenced. At some point, the surgeon glanced at the EKG and noticed severe bradycardia. The ventilator was not restarted by the anesthesiologist. The ventilator is typically paused for 20–60 seconds to prevent motion-induced blurring of the image. The patient expired [43].

Cardiac (heart) surgery typically requires the use of cardiopulmonary bypass (CPB). During CPB, the CPB machine takes over both the pumping function of the heart and the ventilation function of the lung. Therefore, during CPB, the anesthesia machine ventilator is usually not needed, and is turned off to prevent unnecessary ventilation-induced lung movement that can interfere with surgery. During this period, physiological respiratory and circulatory monitors can be turned off or their alarm signals inactivated to prevent nuisance false alarm signals. At the conclusion of the CPB period, the heart resumes pumping blood, and the CPB machine pump is stopped. Lung ventilation must be resumed prior to discontinuation of CPB to prevent circulation of non-oxygenated blood which can cause organ damage. The anesthesia/surgical team has to remember to resume ventilation and manually re-start the anesthesia ventilator. Patient injuries and deaths occur when the team forgets or delays resumption of ventilation. This is a longstanding problem that continues to occur. Immediately following CPB, the heart and other major organs can be especially susceptible to injury from poorly oxygenated blood [43].

A 49-year-old woman underwent an uneventful total abdominal hysterectomy (removal of uterus and cervix through an abdominal incision) and bilateral salpingo-oophorectomy (removal of both sets of fallopian tube and ovary). Post-operatively, the patient complained of severe pain and received intravenous morphine sulfate in small increments. She began receiving a continuous infusion of morphine via a patient controlled analgesia (PCA) pump. A few hours after leaving the PACU [post anesthesia care unit] and arriving on the floor of the general ward, she was found pale with shallow breathing, a faint pulse and pinpoint pupils. The nursing staff called a "code" and the patient was resuscitated and transferred to the intensive care unit on a respirator [ventilator]. Based on family wishes, life support was withdrawn. The patient expired. A case review affirmed death due to PCA overdose [43].

**SOLUTION SPACE – Improved Patient Safety due to an Integrated Clinical Environment**

An elderly patient with end-stage renal failure was administered a standard intravenous (IV) insulin infusion protocol to manage her blood glucose, but no glucose was provided (oral or intravenous). Her blood glucose dropped to 33mg/dl and then rebounded to over 200mg/dl after glucose was administered [43]. The fluctuation is harmful for the patient.

The extreme swings of hypoglycemia and hyperglycemia should be avoided. A patient receiving IV insulin infusion must be continuously monitored for blood glucose level. An integrated platform using a workflow may automatically adjust the IV syringe pump rate to deliver the insulin according to the real-time blood glucose levels provided by the glucose monitor (mobile data dashboard also available in real-time at point of care). Integration of active device data with other patient specific data (perhaps from electronic health/medical records, EHR/EMR) may optimize the physiologically desired patient-centric maintenance of blood glucose values. The patient-specific target range integrates in the decision process the patient's record for weight, target glucose range, typical insulin dosage range (and correction factors, if any), glucose responsiveness to meals (insulin-to-carbohydrate ratio, glucose tolerance test data) and continuous assessment of vital signs (temperature, pulse oximetry, end tidal carbon dioxide, electrocardiograph, blood pressure, respiratory rate).

Hence, an integrated system will host a physiologic closed-loop control (PCLC) system algorithm to monitor and analyze the data from multiple devices as well as relevant health records in order to use an evidence-based (perhaps intelligent) decision tool to regulate the delivery of IV insulin to maintain the blood glucose values within the clinically desired range for this specific patient. To maintain homeostatic glucose levels within the target range, the system can also change/modify the glucose infusion rate using the integrated system-hosted algorithm. The IV insulin and glucose infusion changes can be automated and the system can execute changes to maintain physiological homeostatic balance without human intervention but will issue alarm/alert concomitant with the change (to mobile phone or point of contact nurse/physician) in case the medical situation warrants a medical professional to override the automated decision of the PCLC system algorithm.

The integrated system and algorithm can also trigger a software defined acuity-based ***contextual*** alarm and alert if patient safety was deemed at risk either for engineering reasons (device malfunction, null data, interface error) or medical causes (patient is at high risk for myocardial infarction). Hence, the maintenance of glucose levels and/or resultant physiological issue may reach beyond the design/scope of the PCLC bounded algorithm-driven decision process. The embedded logic may no longer effectively and safely manage the patient's (homeostatic) outcome by adjusting IV insulin dose or glucose infusion [43].

INNOVATION SPACE – The Unstoppable Tsunami of Ubiquitous Digital Connectivity

Making sense of sensors, sensor data and reflected radio frequency waves could lead to life-saving solutions. US reports indicate that nearly half a million patients may be dying each year due to preventable medical errors. One source of error is our lack of knowledge about the physiological status of the patient. Healthcare systems are sluggish in their progress and zeal to synthesize the data in order to provide the necessary information. Taken together, we are swimming in sensors and drowning in data yet slow to integrate the systems on a common platform to alleviate the burden of death due to errors. The common notion that such integration may be prohibitive in terms of cost is not supported by studies which claim that improved patient outcomes can be delivered within acceptable cost [44].

The importance of being frugal in healthcare carries a substantially heavy moral burden but ignoring fiscal reality and affordability may not be prudent [45]. In our haste to be penny-wise, we are often pound-foolish and court disasters. Suggestions in this paper are tainted by our lack of data and scenarios from other parts of the world but it may not be entirely callous to generalize [46] that elderly individuals are often rushed to emergencies due to falls. Accidents and emergencies are hemorrhaging [116] the healthcare system of its resources. Falls lead to hip replacements and brain injuries which are cost intensive [47].

It seems reasonable, therefore, that prevention of falls should be quite high on the list of priority for individuals over 65 years of age who are covered by Medicare and Medicaid in the US. Is it? In the US alone, more than 25,000 deaths [Figure 3] were reported in 2014 due to unintentional falls. The cost of treatment for falls in the US exceeded $31billion in 2015. Millions are treated in emergency rooms and thousands hospitalized, each year [48].

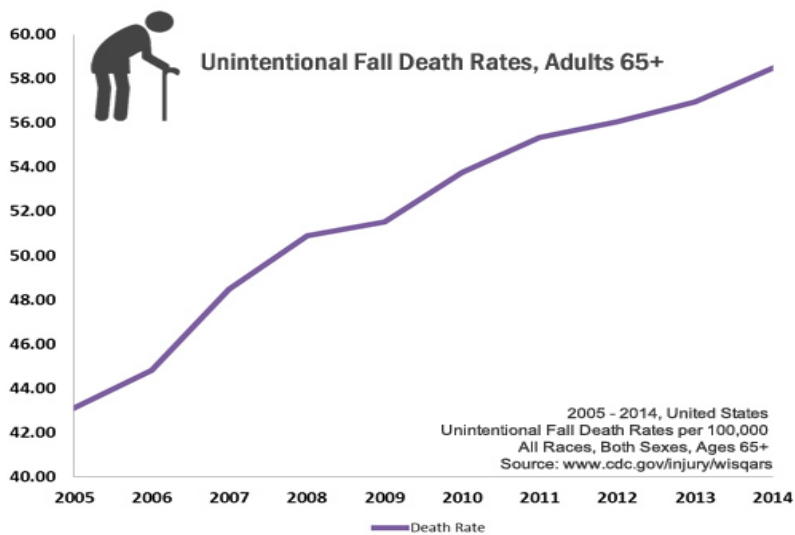

Figure 3: Mortality and morbidity due to unintentional falls may be preventable [48].

Prevention of falls is one example where integration [49] of medicine with innovative technologies [50] offers solutions. The latter is driven by principles of convergence [51] through collaborations between engineering [52] and medicine [53]. Confluence of ideas in advancing healthcare must include the entire spectrum including prevention, primary care, accidents and emergencies, palliative centers, hospices and dying with dignity [54].

Monitoring individuals in homes, hospitals and nursing facilities using reflected (RF) radio waves [55] eliminates the need for sensors. Eliminating the need to wear a tag or sensor or transponder or any object on the body is helpful to elderly people who may be forgetful or suffering from neural afflictions. Analysis of reflected radio waves may predict (Figure 4) if a person is likely to fall. Reducing the time from the instance of the fall to medical attention is key to reducing long term brain damage, hip replacements and fatalities. Today, we have the tools to prevent mortality and morbidity due to unintentional falls, albeit, in part. As soon as an event (fall) occurs, the system can connect with ambulatory services and alert care-givers [56]. Integration of the patient's existing medical data with the patient *en route* to the hospital, may reduce delays in the emergency room and improve quality of care.

The principles and practice of connectivity, when taken together with an integrated clinical environment (ICE), may enable the system to combine device data in hospitals as well as data from home, wearables, edge devices, nutrition indicators and wellness monitors. This is the tsunami of change facing preventive care, healthcare and the medical-industrial complex. Science and engineering [57] are catalyzing tools and technologies to advance medical [58] cyber physical systems [59] and fueling an explosion of the medical internet of things [60] which can save lives and deliver some form of basic healthcare to billions [61].

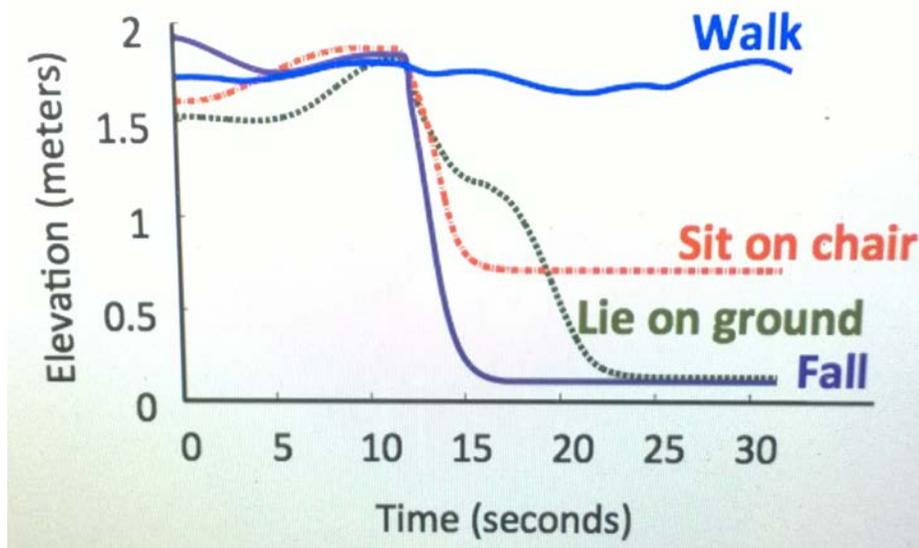

Figure 4: Predict the possibility of fall by analyzing the data from reflected radio waves [55]

**DIGITAL TRANSFORMATION OF HEALTHCARE – Connected Value Chain of Medical IoT**

The grand challenge implicit under the umbrella term medical internet of things is the ability to have an open mind and rise above the restricted view of medical IoT as a new thing or technology. IoT is a digital by design metaphor. By extension, medical IoT is a design metaphor, too. Why IoT is a design metaphor may be better understood if traced back to its roots in a seminal paper [62] and recent publications, in 1991 [63] and 2000 [15]. The design principle is not limited to healthcare or the medical domain. It is applicable to a vast number of domains [64] including manufacturing [65], transport [66] and energy [67].

The digital by design paradigm is made possible due to rapid diffusion of ubiquitous connectivity. The principle of connectivity, when transformed into practice, for example, Integrated Clinical Environment (ICE), is applicable to medicine and healthcare. As defined by ASTM F2761-09 standard [43], ICE is an architectural framework (work in progress) for integrating the clinical environment. It is a step toward enabling interoperability which is key to implementation, not only for medical IoT but anywhere we can apply IoT by design.

The global appeal of IoT as a design metaphor is based on the potential of IoT to spur economic growth [68]. Since hype is the main driver for public relations, the bombastic megalomaniac projections [69] are not predictions but perhaps clues to half-truths.

What appears to be true is the ability of the IoT as a design to reduce transaction cost [70]. The robust principles of transaction cost economics [71] when viewed in the context of IoT appears to support the definitive trajectory of IoT as a conceptual tool for global digital transformation expected to reshuffle prevalent trends and create new markets [72].

Observing the mere mundane fact of comparing prices of goods using your smart phone and then opting to buy the product you choose from the comfort of your home is a form of e-commerce. It is retail IoT as well, because connectivity may allow you to explore your purchase and your trigger to purchase may be due to the embedded IoT functions.

Use of medical IoT in the prevention of falls may reduce a portion of the $31billion direct costs for treating unintentional falls in the US. Research driven entrepreneurial innovation created economic growth [73] and even a modest (10% - 20%) reduction in direct costs will save billions for US tax-payers. Taken together, medical IoT will reduce transaction cost.

Transaction cost in the value chain of ecosystems drives the cost of creating end to end connectivity. Digital solutions are rarely vertical and engage diverse supply chain partners. Digital transformation will morph classical models and supply network strategies. Services which offer data, value, information and knowledge synthesis, are preferred over products.

## MEDICAL IoT – INTEGRATED CLINICAL ENVIRONMENT (ICE) – Information Synthesis

The lack of integration of data in the context of medical devices is a major contributor to lapses in patient safety in the US [74] and worldwide. Lack of granular data from and interoperability between medical devices (attached to a patient in a hospital or from remote monitoring at home) are factors which compromise patient safety. In the absence of appropriate data analysis and synthesis, the ability to produce actionable information relevant to the collective status of the patient must depend on medical professionals to combine the data at the point of care. In this context, integration and interoperability may improve efficiency, safety and security. It is anticipated that it will lower the rates of preventable medical errors and signal a significant improvement in quality of patient care.

The grand challenge of the medical internet of things (MIoT) is to sufficiently enable the deployment of patient-centric and context-aware networked medical systems in all care environments, ranging from hospital floors to operating rooms, intensive care units to home care units. Heterogeneous devices in each care environment would effectively share data (efficiently, safely and securely) to minimize preventable errors often introduced by humans.

Mobility of medical devices between different care environments or from patient to patient presents challenges. While in transit, they must remain secure yet "open" for "discovery" by other devices in much the same way that consumer IoT accessories, for example, Amazon Echo Dot, pairs with your smartphone or the laptop when it is in the cone of connectivity. Once discovered, the devices must self-organize and form the *ad hoc* mesh network of devices, specific to the patient. They need to interoperate with, verify and execute safe, authorized and compliant operational profiles.

This generic *modus operandi* is far from the norm. To transform this simple procedure to reality, we need common or standard architectures which will optimize and prioritize the necessary balance between utility, reliability and safety with those of accuracy, security and privacy. Unless widely adopted, reference architectures, blueprints or standards are sterile, impotent and quite useless [75].

The Integrated Clinical Environment (ICE) framework, as defined by the ASTM F2761-09 standard [43] is a step toward enabling the interoperable MIoT vision. ICE is an architecture for a medical data and device interoperability platform [76]. The debate rages on about open [77] vs partially closed platforms [78]. Platform is a new word for the conceptually similar "common bus" in computers [79]. ICE is a platform where devices may connect via standard interfaces or API's in order to contribute, select or analyze data. Actionable information relevant to the patient can be updated, shared, published or routed. OpenICE is the reference implementation of this platform vision. The medical community may access OpenIce (www.openice.info/ and www.mdpnp.org) and adapt and/or adopt for their use.

Layering ICE with a variety of engines for distributed data acquisition, data curation and data analytics (dynamic composition of engines based on the nature of data), crowd-sourcing open apps to innovate using available (de-identified) data, introducing novel algorithms for non-obvious relationship analysis (correlations) and semantic data dictionaries, are sign posts for future development. We are keen to find unknown and unchartered paths beyond the horizon. Integrating cybersecurity, security, privacy, device discovery, authentication, standards, interoperability and data de-identification [61] are a few of the more immediate tasks. Security for interconnected and dynamically composable medical systems are a priority not only because they are critical to patient safety but also because they are mandated by HIPAA, the Health Insurance Portability and Accountability Act of 1996 [80].

In most environments, physical objects – the medical devices – are likely to be mixed and matched in an *ad hoc* fashion to serve the patient (dynamic system composition). It may be a conceptual stretch but compare it to an assembly of Lego blocks where the patient is the "base" to which the blocks are connected, either directly (device) or indirectly (printer or screen) or remotely (wireless heart rate monitor). In the rush to create a working solution for a patient in an emergency room, medical staff may "grab" what is available at first sight depending on the need (infusion pump, sphygmomanometer, plethysmograph, ventilator).

Did they verify the security mechanisms and authorizations which are configured on these devices? How will they support automatic verification that the system components being used are as intended in the correct clinical context? How will they determine that the components are authentic and authorized for use in that environment? Where can they log in to confirm that the devices are still approved by the hospital's biomedical engineering staff and certified to meet regulatory safety and effectiveness requirements?

In the long term future, as soon as a staff touches a medical equipment or a device, an automated association will be instantiated on her/his mobile device and/or in a control device which will serve as a digital footprint connecting a specific human with a specific object (for example, a baggage handler wearing a RFID and sensor embedded hand glove picks up a piece of luggage tagged with a "smart" label). The association can extend to other humans (in this case, nurses, physicians, patients) or objects (equipment, devices). In this manner, the staff rushing around to gather the devices and connect them to a patient is free to do her/his job while the "digital twin" evolves [81] to capture the patient, team of medical professionals and the devices connected to the patient. Each team member will receive a copy of the "digital twin" on their smart phones. The questions in the preceding paragraph will be digitally queried and unsatisfactory outcomes could trigger alarms or alerts (for example, change device or charge equipment or re-boot EKG machine).

The "digital twin" will capture the network of activities in progress in real-time, monitor for errors, serve as an active virtual assistant (avatar) to the team and store a continuous log of events. Currently, none of the above is available but digital twins [79] are emerging.

Today, as far as medical device communication is concerned, few of the existing or proposed standards, for dynamically composed and interoperable medical devices and information systems, include sufficiently comprehensive or flexible security mechanisms to meet current and future safety needs. There are gaps between required security properties and those that can be fulfilled even by combinations of currently standardized protocols [44]. Safety considerations in these standardization efforts are effectively incomplete due to a lack of appropriate security analysis.

Regulators are also noting the importance of incorporating security for safety and privacy in the medical domain. The FDA is calling for medical device manufacturers to address cyber-security issues for the *entire* lifecycle of the device. It begins with the pre-market stage [82] from initial design through deployment and end-of-life or post-market stage [83]. FDA may use these draft guidelines as a basis for clearing medical device submissions [84].

Imminent fears of regulatory hurdles, associated with security, further discourages device manufacturers from delving deeper into installing security which may be subject to approval by the FDA [85] for US markets. Elsewhere, the considerations may be different.

The device-centric view of security as a company-specific endeavor may be the primary barrier in this traditional hospital-focused effort which excludes the distributed network-centric view which is rapidly evolving [86]. Cybersecurity, security and privacy related to any device and its data is a patient-centric issue and no longer a company or device-centric issue. Each device and its data is critical yet one part of a larger, distributed, temporal aggregate of a network of devices, data, decisions, events and records, linked over time, to a person and/or patient. It is a time series of events. The demand for total security and pervasive cybersecurity to ensure that privacy is not punctuated remains to be analyzed.

Solutions may evolve from alliances, organizations or groups working as a team bounded by goals which must rise above the mantra of amplifying shareholder values and take into account social responsibility to promote stakeholder values. The question without a good answer is the amount of security and privacy which is enough or is sufficient or is needed. Pervasive security and cybersecurity may be an illusion and costs may be truly prohibitive. Funding for pervasive cybersecurity may kill the patient by draining away the resources necessary for medical purposes. Perhaps passengers and drivers in body armors may have a lower risk of injury or fatality in case of an automobile accident but as a society we have decided to limit our sense of safety and precautions to safety belts and air bags in vehicles.

An analogy of punctuated security and privacy is the use of a tire pressure sensor that you bought online to monitor your car's tire pressure using an app on your iPhone. A hacker went into the control system of your BMW using the external (not factory installed) tire pressure sensor and disabled your brakes. The shield of cybersecurity was punctuated. Similar events using other IoT devices were predicted in June 2015 [87] and massive attacks were observed on 21 October 2016 [88]. Healthcare is not immune from such attacks [89].

Healthcare is no longer bound by the traditional medical institution and it is no longer in complete control of the flow of health information. Hence, security considerations must be regarded and regulated as public goods. The new and fractured state of health, wellness and emergencies are distributed problems. It cannot be solved by hospitals, alone.

How much of which data must be protected or should be kept private will change in a dynamic manner depending on personal preferences and health coverage in different nations. Travel and tourism makes it imperative that humans receive treatment when outside their countries.

Taken together, issues of interoperability, dynamic composability, variant configuration and levels of operational flexibility are all important. These factors call for different thinking about principles, infrastructures and tools which can be globally deployable yet subject to local optimizations [90].

However, no single solution can act as a panacea for all scenarios and applications. We need to balance new approaches with the old [91] in an effort to converge healthy idealism with doses of pragmatism. Fear of short term vulnerabilities may, justifiably, evoke resistance to reduce the dead weight of old technologies even though pockets of conviction exist about the value of distributed agent-based approaches [85] as a robust long term prescription for digital transformation.

The ICE framework, as defined by the ASTM F2761-09 standard [43] provides an approach for integrating heterogeneous medical devices and coordinating their activities to automate clinical workflows. Medical devices that conform to the ICE standard, either natively or using after-market adapters, may interoperate with other ICE-compliant devices regardless of their manufacturing origin.

A similar paradigm exists in the computing and communications domain, leading to an explosion of devices supporting WiFi, USB, Bluetooth and other telco standards. It may be useful to think of ICE architecture, ICE standard and OpenICE reference implementation as foundational as the http or hypertext transfer protocol [92]. The global impact of the non-proprietary, open standards based http, needs little over-emphasis. The mission of ICE is similar to that of http for medical equipment, devices and any object used for healthcare.

ICE standard will enable dramatic improvements to patient safety because cross-vendor inter-device communications may significantly reduce preventable medical errors. Examples include patient transfers from the Operating Room (OR) to Intensive Care Units (ICU) and reducing false alarms or deaths due to Patient-Controlled Analgesia (PCA). In both of these examples, synthesis of data from diverse range of medical devices may enable the generation of real-time actionable information from the system to a point of care physician/nurse to prevent medical errors and reduce patient mortality and morbidity.

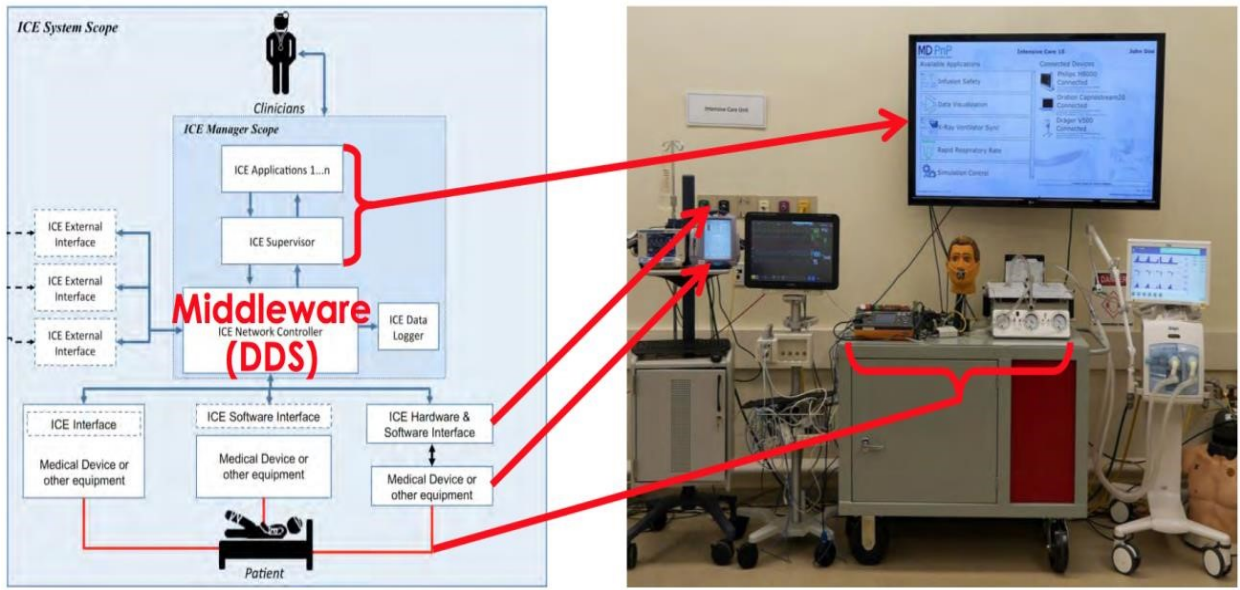

Figure 5. General architecture of ICE and an instantiation of it in a test setup at MD PnP Lab

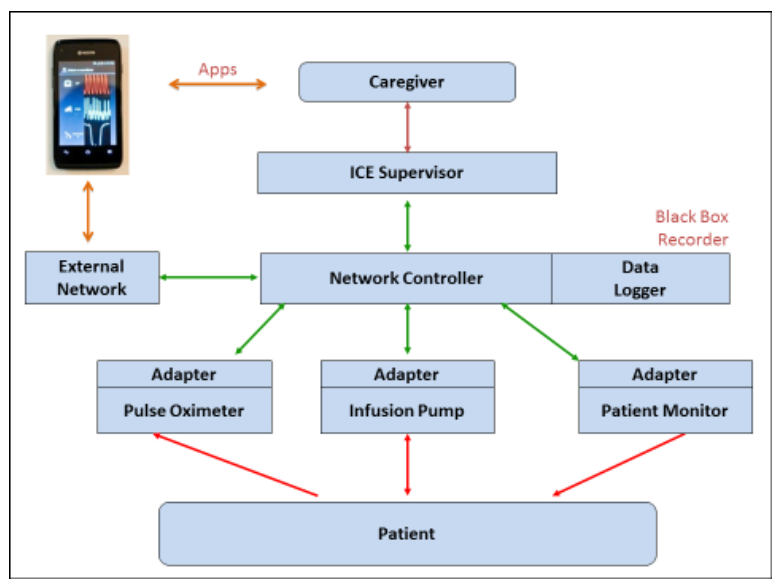

Figure 6: Overview of the Components of the ICE System

The *ICE Network Controller* is a high-assurance middleware that forwards data or commands to or from ICE applications and devices, ensures communication related quality-of-service and is agnostic as to the intended use of the clinical apps that it may support. It also manages the **discovery** and connection protocol for devices that wish to connect to the system.

In view of its central role in ICE communications, it is imperative to install high-performance and context-aware security support in the ICE network controller. The major functional security requirements for the ICE network controller are:

[a] authentication mechanisms for validating the identity of devices and apps, vouching for their provenance and ICE compliance,
[b] dynamic and flexible yet user friendly mechanisms for defining and enforcing access control policies for various ICE configurations in different environments,
[c] mechanism for secure device and app discovery,
[d] secure auditing mechanism and
[e] tools to guarantee the integrity, freshness (perishability) and confidentiality of data.

Functional requirements should minimize negative impact on non-functional requirements such as performance, availability, robustness and ease of use for clinicians and developers.

Safety considerations in standardization efforts are incomplete due to a lack of appropriate security analysis. To address this, we developed a prototype of ICE based on DDS or Data Distribution Service middleware [93] as the ICE Network Controller (please see Figure 5).

DDS is a communications API and an interoperability standard that provides a data-centric publish-subscribe model for integrating loosely coupled real-time distributed systems. DDS is data-centric. DDS separates state [94] management and data distribution from application logic and supports discoverable data models. This exposes data model to communication middleware, enabling DDS middleware to reason and optimize the performance of data movement within the system. To customize run-time behavior and achieve desired performance profile, DDS allows publishing and subscribing entities to express several quality-of-service (QoS) parameters. The offered versus requested QoS requirements of the participating entities are matched before any communication can proceed.

The *ICE Supervisor* provides separation/isolation-kernel-like data partitioning and time partitioning. It makes sure that the information cannot inadvertently leak between apps and apps cannot inadvertently interfere with one another. It provides real-time scheduling guarantees that the computation in one app cannot cause the performance of another to degrade or fail. It also provides a console that allows a clinician to launch apps, monitor their progress and provide user input during app execution. The ICE Network Controller and Supervisor may be incorporated together and deployed as a standalone ICE Manager.

*ICE Applications* are programs that accomplish a clinical objective by interacting with one or more devices attached to the network controller. As each app executes in the supervisor, it defines the intended use of the current ICE configuration. ICE medical devices never interact directly with each other (for safety and security reasons). All interactions are coordinated and controlled via the ICE apps. It is crucial that ICE apps exactly correspond to the specified task for which they were designed.

*ICE Equipment Interfaces* declares the functional capabilities of the device (for example, format of its data streams, commands to which it responds) along with non-functional properties of the data such as the rate at which data elements are streamed from the device. It is crucial that ICE Interfaces are designed with considerations for usable security, for developers and clinical end-users.

*ICE Data Logger* is dedicated to logging communication and other important events within the ICE Network Controller and ICE Supervisor. Data logger should also record all security related events. The data logger can be used by medical professionals to re-create the events and actions.

It may also serve to protect against institutional vilification and indemnify the professionals in case of charges of death due to malpractice or insurance claims/payments. Physiological networks can interact and cross-react in many unpredictable ways which could lead to a cascade of deleterious events without human intervention (human is not a fault). In a very litigious society, as in the US, such incidents are often followed by "see you in court" tweets. Thus, the burden is on the medical domain to provide verifiable evidence of the sequence of events and identify each human associated with each stage of that process. Without the ICE data logger, this becomes a network reconstruction problem and network reconstruction remains an outstanding challenge [95] even in the absence of physiological complexity. ICE data logger can offer time-stamped series of the sequence of events which may exonerate the medical institution and professionals, if no human / preventable error was involved.

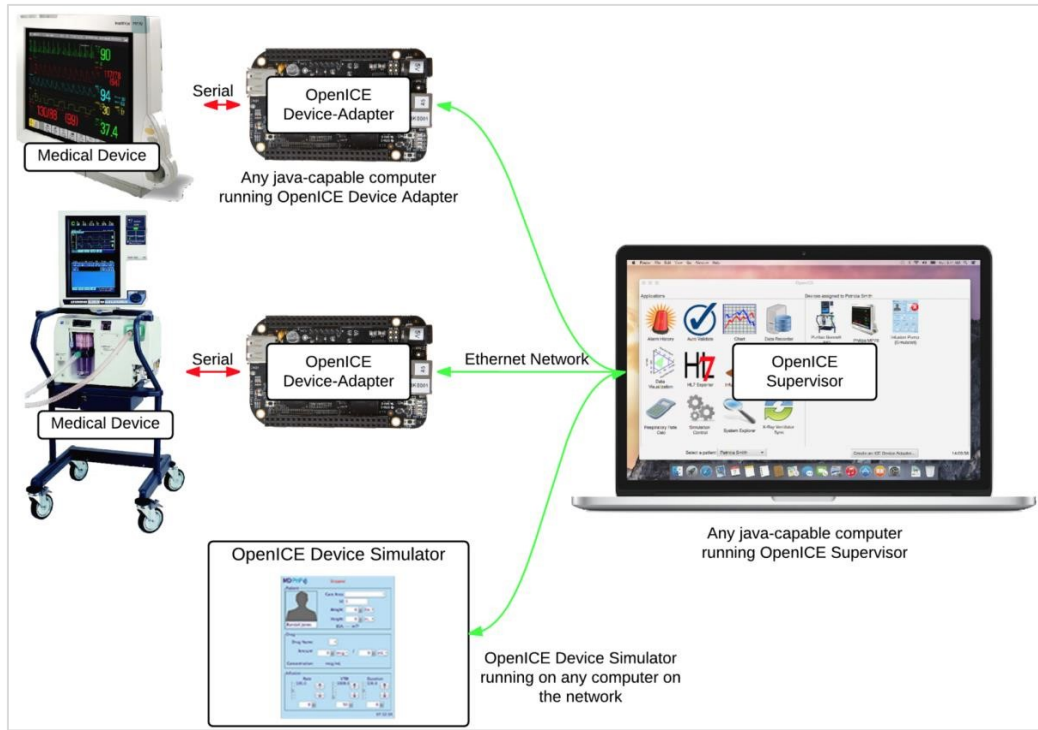

Figure 7. OpenICE from MD PnP Lab enables connectivity between various types of devices.

Standards-based medical device interoperability can also provide real-time comprehensive population of a patient's Electronic Health/Medical Record (EHR/EMR). It will permit the creation of integrated error-resistant medical systems and support advanced capabilities such as (a) automated System readiness assessment, (b) physiologic closed loop control of medication delivery, ventilation and fluid delivery, (c) decision support using machine learning and/or rudimentary artificial intelligence applications, (d) safety interlocks, (e) monitoring of device performance, (f) plug-and-play modularity to support "hot swapping" of replacement devices, (g) selection of "best of breed" components from competitive sources and (h) innovations to improve patient safety, treatment efficacy and workflow efficiency.

The vision of data synthesis and data interoperability depends on adoption of these (ICE) standards by healthcare delivery organizations and medical device manufacturers as well as endorsement by organizations (eg AMA) and regulatory bodies (eg FDA). Medical Device "Free Interoperability Requirements for the Enterprise" or MD FIRE [96] is a guide to coalesce the ecosystem necessary to create a critical mass which may initiate the steps to reduce preventable medical errors. The pursuit of ICE as a standard and interoperability between other ICE-like global standards are essential. ICE advocacy by organizations and adoption of ICE by choice or regulation will enhance patient protection and patient safety.

**BEYOND ERRORS – The Opportunity and The Need**

The problem of preventable medical errors is, at least in part, created by the penchant for profit. Proprietary semantic data dictionaries, locked data interfaces and upselling services are tools of the trade used by the medical-industrial complex. Adoption of open standards and open platforms [97] may partially ameliorate this dysfunction. But, the corporate need to monetize is also necessary to spur economic growth and incentivize innovation. The balance between idealism and pragmatism is crucial for our path to (ethical) profitability.

The far bigger opportunity in seeking better quality of EHR/EMR is not limited to potential for integration with data from devices and home monitors or inclusion of wearables. There are other sources of data in a fractured state or uncollected. Data from transcriptomics, genomics, proteomics, metabolomics, imaging, and pathology may reveal patterns or interplay between morphogens and morphogenesis or factors that influence immune responses [98] or eliminate pathogens [99] or analytes which predict heart failure [100].

This vast resource of data, if collected, combined and analyzed (if made available from an open repository) will aid systems biology to uncover clues cryptic in networks of proteins or genetic circuits or cellular signals. These signals connect, converge and collude in a myriad of ways to promote health or cause disease. Elucidating the elusive paths taken by these target molecules may be one part of the future progress of precision medicine [101]. The toilet in our bathroom may become our best ally [61] in preventive medicine.

Transforming these target molecules to prognostic tools may open the Pandora's Box on population genomics and populations at risk. Global medicine must find ways to address health and healthcare issues for billions. Moving from expensive micro-arrays to very low cost nano-sensor [102] arrays or nano-wires [103] to detect single molecules may hold promise for diagnostics. However, the ability to detect is only a part of the equation. Reusability is a key driver to reduce cost. Analyte association is as important as dissociation to replenish and recycle the detection tool. The time, duration and circadian rhythm of the signal is pregnant with information. An "always-on" transmission is required to capture any signal, anytime. The form factor and stable vs unstable location (topical, epidermal, subdermal, subcutaneous, ingestible) will influence efficacy, accuracy and adoption. Invention [104] and innovation of new technologies [105] may be necessary to detect, transmit and capture signals *in vivo*.

Data leading to diagnostic tools and data logging of transmitted signals to identify patterns will need a variety of tools as well as new thinking and re-visiting old ideas. We will need data curation, pattern analysis, non-obvious relations, chemistry, engineering, computer science, nanotechnology, telecommunications, software, system of systems and medical cyber physical systems. Confluence of transdisciplinary knowledge domains will be essential to synthesize the outcomes and coordinate the compass with the road map.

This is yet another dimension of the anticipated digital transformation in medicine. Medical IoT is a part of this forthcoming tsunami from ubiquitous connectivity. The confluence of molecular connectivity with 3D/4D printing and distributed additive manufacturing-on-demand facilities may deliver medical tools to benefit billions. The new business of billions may flourish if rooted in the concept of micro-payments for services.

Current cost of hip replacement in India is about US$5,000 but at a per capita annual income of $1500, nominal ($5350 by purchasing power parity), it is still out of reach for many Indians (even though the cost is about one tenth of that in US). If this product was transformed into a service and distributed over the lifecycle of the implant (say 10 years) then the daily micro-payment amounts to $1.40 (or more depending on cost of service). For 5.6 billion people in the world surviving on less than $10 per day, the amount spent on healthcare will be minimal. Can the demographics, who need the hip replacement the most, afford to bear the cost of $1.40 per day for 10 years if they earn about $10 per day?

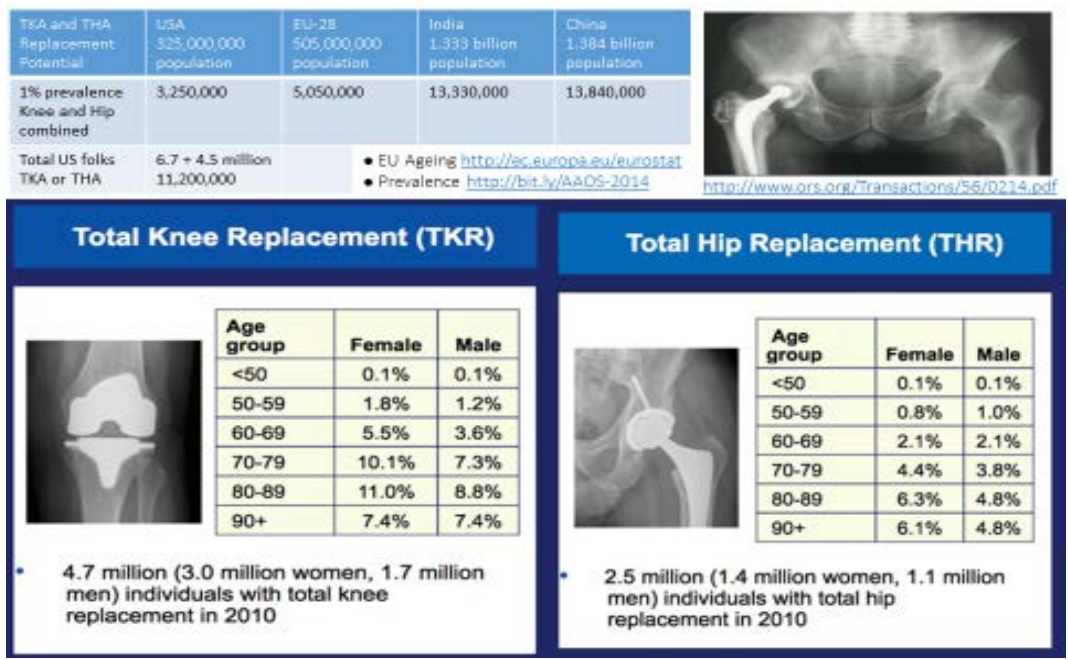

Table 2: If 1% of the population of India and China could afford hip or knee replacement then the demand for hip & knee implants (including US & EU) may exceed 35 million [106]

Advances in material science, chemistry of metal additive printing and lowering the capital expenses, related to the manufacturing infrastructure, may converge to lower the cost of the implant. Surgical costs and post-operative costs are, however, significant factors which may not be easily reduced. Spinal implants may need teams of orthopedic surgeons as well as neuro-surgeons. Patient-specific 3D printed implants will need access to imaging tools (MRI), design labs and technicians to produce the implant with a patient-specific fit.

Using an array of sensors to derive data from these implants (for example, for hip joints, femoroacetabular impingement) opens up yet another horizon. To go one step further, we can create a cavity within the implant and use the implant as a Trojan Horse to deliver a variety of molecular tools packed in that cavity. The challenge is to power the tools inside the implant for the lifecycle of the implant and communicate or extract digital signatures, data and signals from inside the implant. The external surface of the implant may serve as an internal "skin" embedded with nano-wire sensors. These nano-wires may serve as docking zones for bio-markers [107] or other analytes, which, when bound, elicit signals. Similar set of issues are applicable in identifying the signal over noise, replenishing the nano-sensor, capturing and transmitting the signal. In addition, the impact of angiogenesis and vascularization on the function of these nano-sensors remains to be determined.

Capturing signals released by our bodies may generate a wealth of data, for example, the temperature of our body (form of radiation/convection). Neural activity inside our bodies generate data we generally do not capture or analyze. This data may offer information about the state of neuromuscular or neurodegenerative diseases, such as ALS [108].

The field of terahertz radiation and imaging is yet another source of data which remains unexplored. Protein electrodynamics [109] suggests that proteins act as biological radios [110]. Proteins emit and absorb terahertz (THz) radiation [111]. Hence, protein signatures may be dynamic or may change, theoretically, depending on the stage of synthesis, modification or mutation. Hence, a potential for comparative analysis of protein signatures to identify modifications or mutations of protein structure. These changes may be implicated in the etiology of disease(s) or dysfunction. Detection of such signals may alter diagnosis and prognosis of disease states, many of which are not sufficiently expressed unless they reach a certain threshold or activation point. Problems pertaining to signal vs noise appears to corrupt data from terahertz imaging [112]. One wonders if innovative application of error correction techniques from information theory [113] or econometrics [114] or some combination of both (or modification of these principles) may help in curating data from terahertz imaging to identify authentic signals released by proteins.

Bringing together data from a variety of domains may revolutionize medicine and trigger thoughts that we don't know how to think, yet. These suggestions may not be correct.

It is to "dare to propose" ideas [115] and re-visit "dots" or renew thoughts which may be waiting to be connected or disconnected.

**THE PRAGMATIC HORIZON – The Influence of Digital Transformation on Retail Healthcare**

The tsunami of the principles and practice of connectivity is expected to usher in an unprecedented era of healthcare information technology that shall be woven into the daily fabric of our lives almost through our entire life-cycle, from conception to the grave.

Healthcare-associated infections may benefit from connectivity [Figure 8] and the connectivity as a medical IoT design principles, if used, may save lives. About 1 in 25 patients gets an infection each year while receiving medical care in the US [116]. Estimated 25,000 deaths per year from about 1 million infections each year from US healthcare at a projected cost of $30 billion per annum. In addition to logging prescribed antibiotics and their use [117] it may be necessary to track and trace individual hand-washing habits and places, beds, patients, equipment they may touch. The tools for such visibility include RFID tracking of objects and their association with events. When added to the power of mobility, this information can be available in real-time and potential nodes of infection predicted or identified before the spread commences or gains momentum.

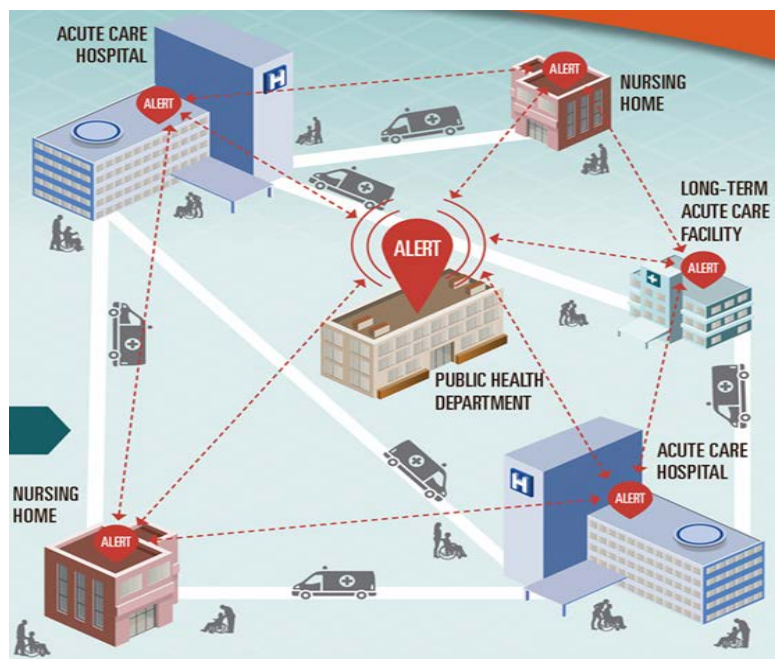

Figure 8: The principles and practice of connectivity coupled with RFID and medical IoT may help in tracking, predicting and containing Healthcare-associated Infections [118]

Digital by design is the fabric integrated above [Figure 8]. Businesses may use similar fabric in an era where IoT may be the dominant design metaphor. Digital entrepreneurship and intrapreneurship is exactly what the doctor ordered. Trans-disciplinary convergence of medicine and engineering may morph the brick and mortar health departments, clinics and pharmacies from its emergency or retail outlet concept. An integrated function for health and healthcare is necessary to decrease demand for high acuity events and the cost.

IoT is poised to recalibrate almost every facet of the business world by taking advantage of the progress of connectivity. Medical IoT will influence health and non-emergency health, as well as general wellness, by connecting precision metabolomics within primary care environments fostering the practice of preventive medicine and clinical attention/action.

Retail clinics and "pharmacies" will undergo transformation to create the 22nd Century service centers for medicine, perhaps something akin to "Jiffy Lube" (Boots, Walgreens, CVS) compared to a visit to Sears Auto Center (hospital). $0.99 Dexa scan, $1.99 PSA tests and $2.99 mammograms may be done in-store when shopping for milk, bread and eggs.

 The transformation will be catalyzed by pioneers who will usher in, albeit in phases, convergence of a wide variety of precision medicine tools applicable on a massive scale and harvest metabolomics data from device-agnostic, protocol-agnostic, platform aggregators (for example, ICE) which will connect to streaming data inside and outside the body (humans, animals). Predictive analytics from person-specific data will be the digital path for clinical "sense and response" system and offer prescriptive analytics. This may serve many facets of preventive medicine, non-emergency medicine but may exclude sudden trauma and ambulatory scenarios.

Retail healthcare may serve as the future point of contact for the confluence of preventive medicine, precision medicine, primary care, tele-health and remote diagnostics. Retail health industries must reform their mission from selling drugs to acquiring data, analyzing and advocating in addition to building alliances to serve individuals who are not patients. The potential of digital by design health IoT will generate business growth and generate massive revenues through pay-per-use micro-revenue schemes. It may help those in the US who are less fortunate [119] and reduce the barrier to entry in L-26 countries [120] where health spending is less than US$50 per year per person for more than 2 billion people [121].

Imagination, invention and innovation must be coupled with wireless telecommunication based remote monitoring where changes in physiological status or alerts could trigger applications via intelligent agents using functional mesh (networks) for multi-directional multi-cast communication of data, information, analytics, intelligence and streams for real time decision support or at-home care or ambulatory access depending on the "sense and response" system of systems that provide one-on-one guidance at point of contact (POC).

The retail health industry must demonstrate this concept on a large scale, to build credibility. It must create the local and global ecosystem of competencies necessary to provide the end to end value chain. It must be driven by the principles of ethical profitability and in practice adopt a micro-payments model for pay-per-event services.

Cybersecurity [122], trust, authorization, validation, privacy, policy [123], regulatory compliance and authentication may require digital ledgers, for example, blockchain-like concepts, to track, trace and secure every instance and events related to every process and nested sub-process. Often data may have long shelf-lives (perishability of medical data).

The complexity calls for a global surge of and focus on, collective entrepreneurial as well as intra-preneurial recombinant innovation. It will create new lines of business and immense economic growth but not through traditional channels and existing business models or organizational *status quo*.

This calls for a new organizational platform approach where credible groups lead and coalesce tools from a diverse array of providers and champion a new form of delivery.

The leadership must embody the relentless pursuit of frontiers without the fear of failure to lift the future plight of humanity through distributed medical care beyond boundaries. One must continuously re-invent to re-align with new research, new inventions, new theories, new ideas, new science, new ways to help people and new customers, locally and globally. If one thinks that any one solution or company or provider or nation holds the key then one may be suffering from disequilibrium [124] due to an incurable ailment commonly referred to (in the medical jargon) as solipsistic bliss.

EPILOGUE – Prologue to Nonlinearity, Known Unknowns and Unknown Unknowns

A systems approach to preventing errors by enabling medical device interoperability depends on data synthesis from the "network" of devices. It may prevent deaths by treating the whole patient rather than responding to separate signals. Insufficient strides [125] and displaced focus [126] are certainly not in short supply [127].

One proposal in this article is to gather device data and acquire data from other sources (genomics, microbiomes, terahertz radiation) in order to combine, synthesize and analyze. The outcome may reveal one or more networks, connectivity, patterns and relationships, both obvious and non-obvious [Figure 11].

Patterns are ubiquitous - from Fibonacci in nature [128] to galaxies [129], even if it is amorphous to the naked eye in very large systems [130].

Network(s) thus revealed may be super-imposed (sub-imposed) on other networks in systems biology – network of gene circuits [131], network of cellular signals [132], neural network of networks [133]. Network layers are part of nature [left, Figure 8]. Cascade of layers and/or networks underlie many functions, such as the OSI model and TCP/IP which is at the core of the internet [center panel, Figure 8]. System of systems may connect and orchestrate public services for towns and communities [right panel, Figure 8] aspiring to join the global league of smart cities [134].

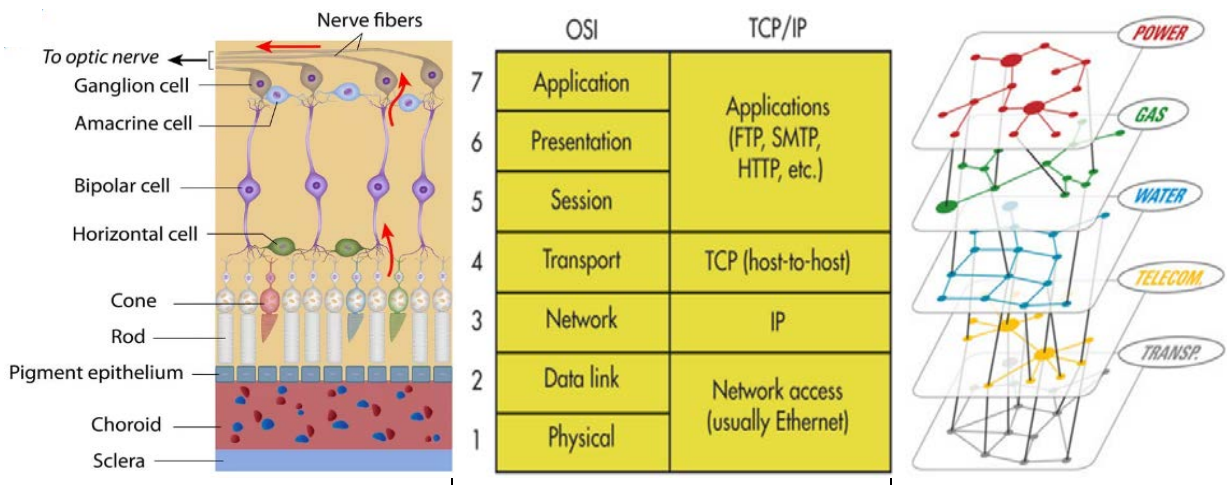

Figure 9: Layers, Networks and Cascades - backbone of the systems approach to function.

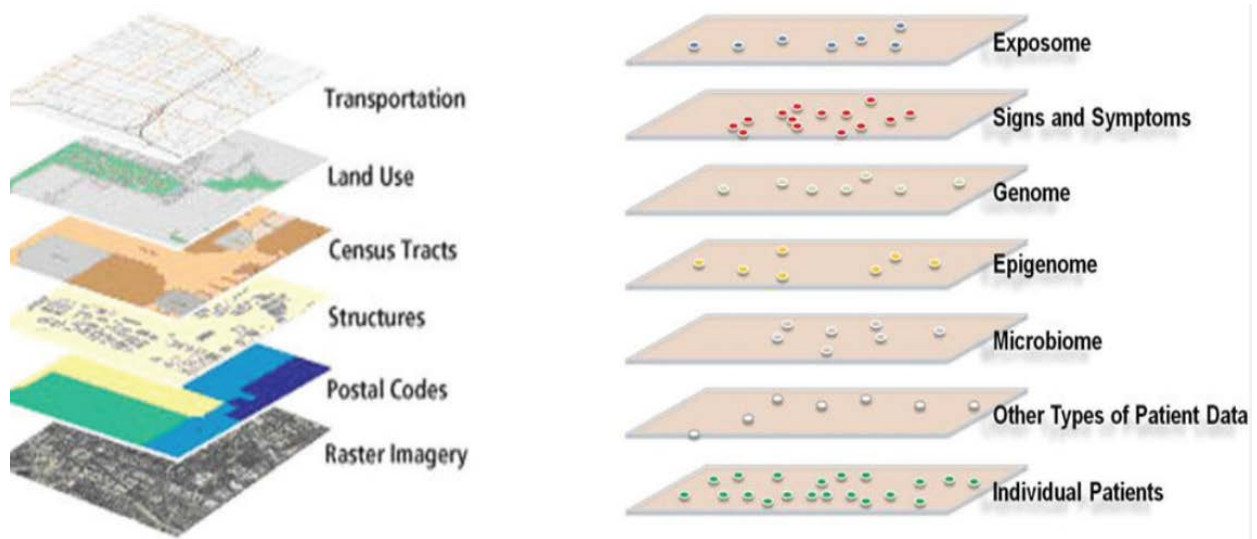

Figure 10: The principle of GIS (Google Map) helped to organize patient-centric data layers

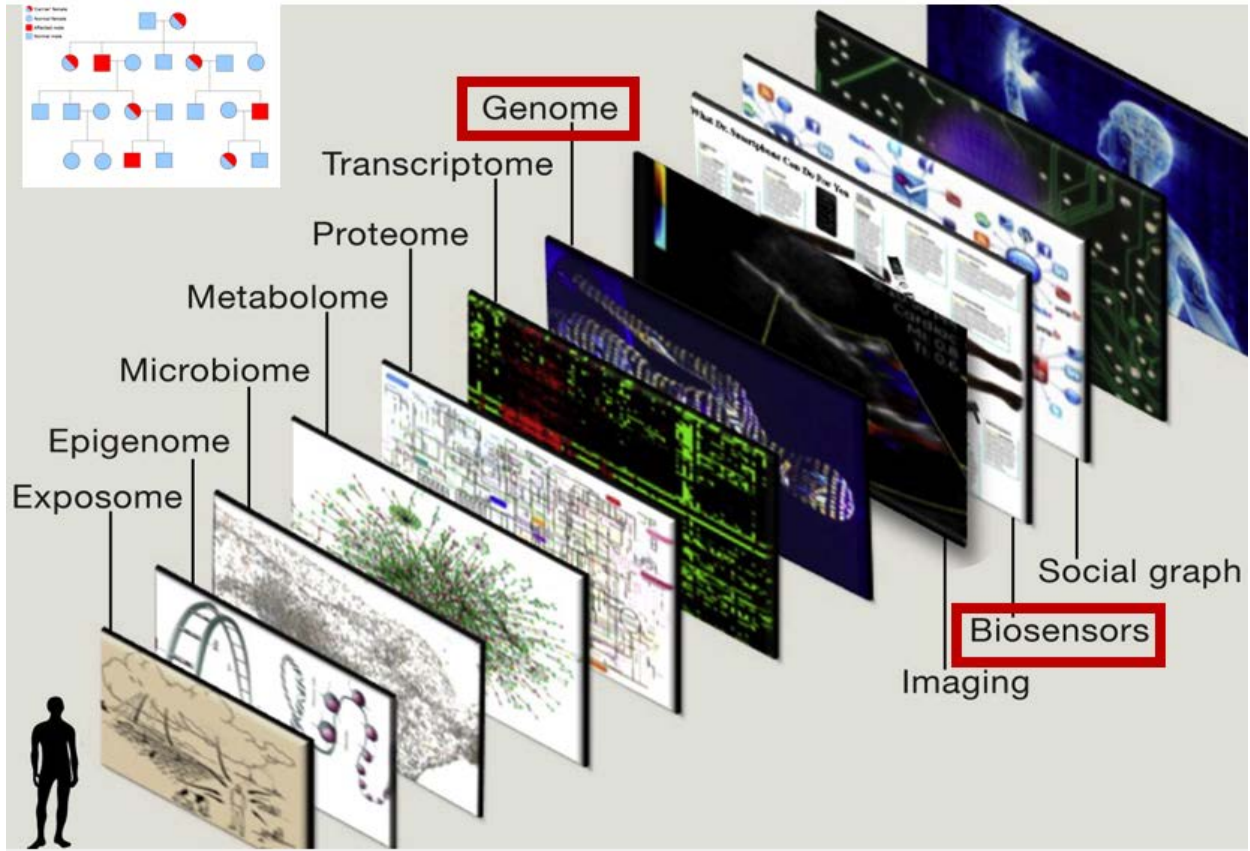

Figure 11: Cascade as a Platform: Combine information from omics with sensor data [135]

Network analytics of such diverse data may need new tools for data curation, as a first step. The relationships within and between these networks may need new description. Some are likely to be linear, perhaps a few of the IFTTT type [136] but non-linearity may be the norm.

Linear associations may have failed to extract the link between gum disease and arthritis [137]. Common gingivitis is implicated in the etiopathogenesis of autoimmune rheumatoid arthritis (RA) due to molecular mimicry. Autoantibody production is triggered by epitope spreading. *Porphyromonas gingivalis*, found in the oral cavity is responsible for certain forms of periodontal diseases. The α-enolase from *P. gingivalis* and humans share 82% homology at the 17-amino acid stretch of an immune-dominant region. Thus, antibodies against *P. gingivalis* α-enolase can recognize the homologous human α-enolase and promote the production of anti-human α-enolase autoantibodies. Anti-citrullinated human α-enolase antibodies were correlated with antibodies to *P. gingivalis* α-enolase in patients with rheumatoid arthritis [138].

Relationships likely to evolve out of this data and analytics effort may influence the future and economics of nations with respect to cost of healthcare.

Treatment of gingivitis using tooth floss, good brushing technique, suitable toothpaste and common sense applications of oral hygiene coupled with low cost adjunctive antibiotic treatment [139] may result in favorable response at the primary care stage. Reducing the risk of autoimmune rheumatoid arthritis reduces healthcare transaction cost, reduces cost of prescription medicine, reduces loss of wages due to disability, reduces morbidity, reduces arthritic pain and improves the quality of life as well as individual productivity which contributes to the national GDP. Ignorance is not a strength given the astronomical difference in cost of prevention due to treatment at the primary care level versus in-patient or ambulatory care.

CDC estimates [140] that about half of US adults (64.7 million American adults) have mild, moderate or severe periodontitis. In adults 65 and older, prevalence rates increase to 70% or higher [141]. Rheumatoid arthritis (RA) affects 1% (70 million) of the world population.

What if we could nearly eliminate gingivitis related autoimmune rheumatoid arthritis?

The 17 amino acid immune-dominant sequence may be subjected to genome surgery [142] using CRISPR/Cas9 tools [143] in eukaryotic systems [144] to edit the coding region (exon) of the human α-enolase gene in a manner which may alter the triplet code for amino acids in the immune-dominant stretch but the amino acid substitution will be highly conservative with respect to the human α-enolase secondary and tertiary structure. It appears that the function of the human α-enolase protein may be refractory to minor modifications [145].

The cost and impact on global mortality and morbidity from autoimmune diseases is staggering. In a few cases, the etiopathogenesis is known and the target is defined. Imagine if gene editing techniques (Cre-lox, ZFNs, TALENs, CRISPR/Cas9) could help to reduce the number of people affected by autoimmune and other diseases. The tools are at hand.

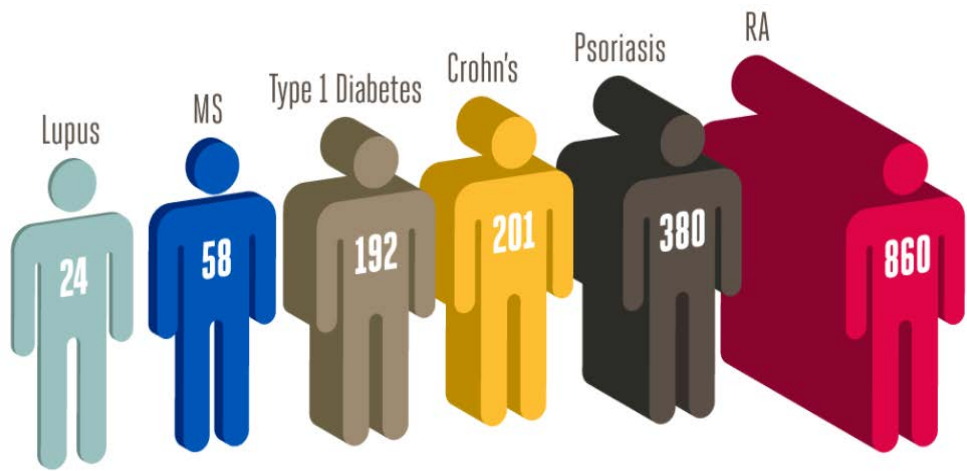

Figure 12: Selected autoimmune diseases and number of cases per 100,000 people [146]

Detection of polyps in the GI tract using disposable ingestible cameras coupled with non-surgical removal of polyps using biomarker-doped excretable micro- or nano-drones [147] may make a difference between an apparently normal life and a life with cancer surgery. The infrastructure for human-mediated surgery or laparoscopic robo-surgery may not be affordable, yet, in many parts of the world. Ingestible cameras and nano-drones may be available as FMCG retail goods, drone-delivered by Amazon or ordered via Alibaba or found in Wal-Mart or in the corner convenience store (where you also swap your solid metallic hydrogen [148] USB battery stick for your electric vehicle's source of energy [72]).

When combined with 3D/4D printing, we may print drones-on-demand (locally) for more specific needs, for example, precision delivery of biopharmaceuticals (3D printed insulin from peptides instead of Humulin or recombinant insulin). These drones may be powered by stomach acid [149], navigated by external wireless signals and carry sensors for analytes or metabolites or bio-markers to update decision systems with data from inside the body.

The breast cancer detecting bra [150], data from the toilet bowl [151] and non-invasive mapping of single nucleotide polymorphisms (coding or non-coding SNP) associated with disease markers [152] may offer clues to improving healthcare and reducing the cost.

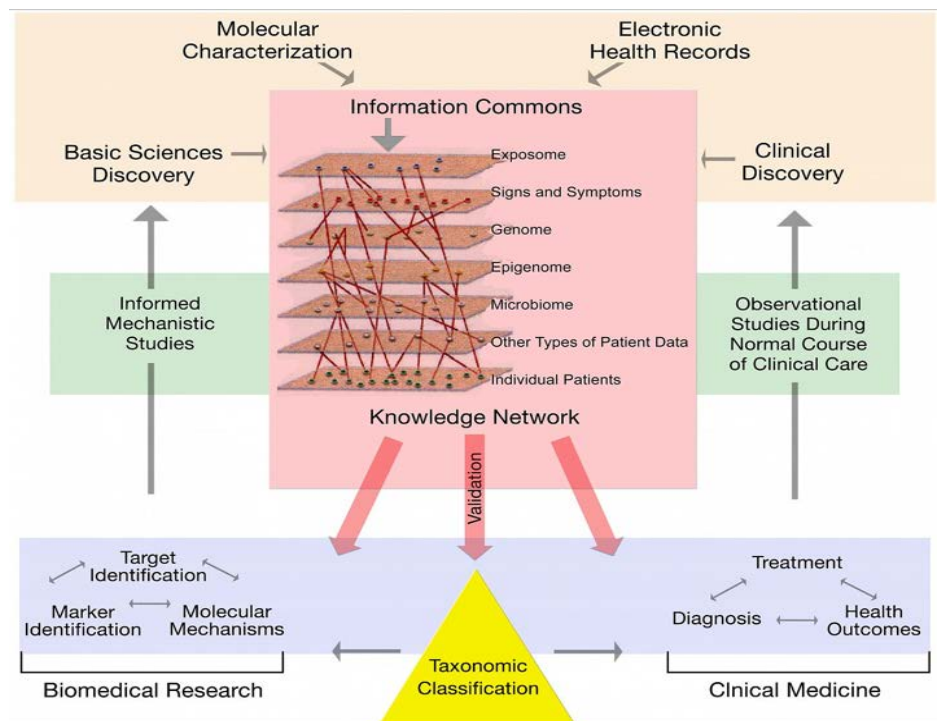

Figure 13: Grand Convergence – Integration of Distributed Data and Information [153]

Making sense of this grand convergence needs new types of data gathering tools [154] and a combination of new data analysis tools with classical tools, such as, machine learning. Elements of rudimentary artificial intelligence [155] in higher order reasoning apps may be useful (claims of "intelligence" in AI [156] are generally due to commercial exaggeration).

Vast numbers and types solvers or engines must be a part of this "analytics" portfolio. The choice of the solver may be triggered by the event or data stream or context awareness or semantic relationship. The architecture of such diversely distributed analytical complexity is uncommon in the current enterprise architecture world. These engines may operate at the edge [157], some in the mist, others in the fog and at the back-end (cloud). Efficacy will be determined by the limits of tolerable latency and time sensitive networking demands.

Crowd sourcing of solvers/engines may enable apps to embed the best in precision analytics. For example, a person in Timbuktu may develop an algorithm which can combine data from blood pressure with levels of uric acid, creatine and creatinine to predict with accuracy the onset of glomerular nephritis which can compromise renal function (kidney). The next gen architecture will have mechanisms to accept, verify and incorporate that app if approved by credible groups [158] or institutions where industry may invest [159] in order to partner with experts [160] who may create or evaluate the tools.

## TEMPORARY CONCLUSION

Neither the US nor the nations aspiring to mimic US-style healthcare can afford to spend at current rates [Figure 14]. Where applicable, digital transformation, if enabled, may partially reduce transaction costs. Improving quality of care and increased access to healthcare for the masses may be facilitated by including digital health and medical internet of things in the portfolio of healthcare. But, these problems cannot be solved with yesterday's tools. Emerging tools, systems and standards must be part of the solution.

Convergence of new thinking, new technologies with new sources of data requiring new platforms, may be essential. To create this new momentum, we need ecosystems which can be formed under leadership which is both credible and can offer guidance for collaboration between different domains. The momentum must be channeled to solve complex problems which are also large, multi-national and the tsunami of data may influence evidence-based policies which may help to lift many boats, not just a few yachts. The sense of purpose of this endeavor must be embedded in a greater sense of the future.

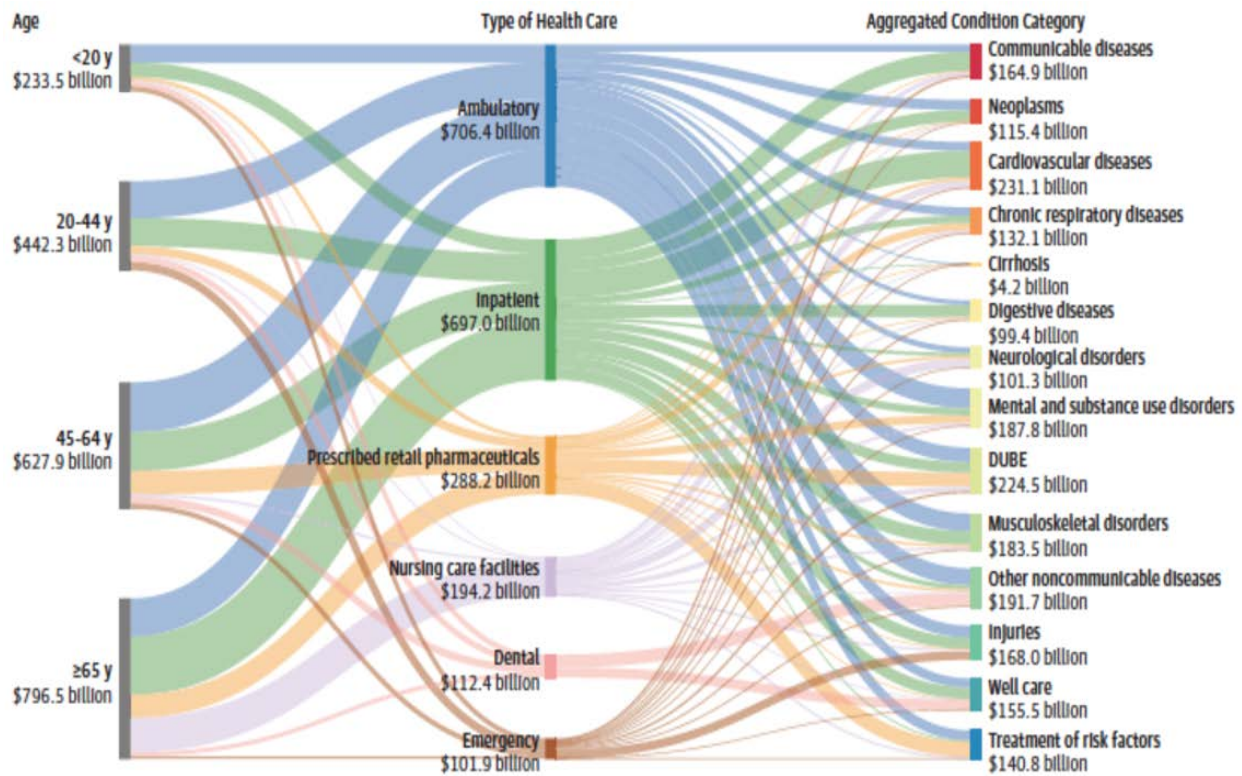

Figure 14: US Personal Healthcare Spending (2013) by Age and Condition Category [161]